\pdfoutput=1
\pdfoutput=1
\pdfoutput=1
\pdfoutput=1
\pdfoutput=1
\documentclass[10pt, paper=a4, UKenglish]{article}
\usepackage{graphicx}
\usepackage{placeins}
\def\Title#1{\begin{center} {\Large #1 } \end{center}}
\def\Author#1{\begin{center}{ \sc #1} \end{center}}
\def\Address#1{\begin{center}{ \it #1} \end{center}}

\newcommand\pubblock{\rightline{\begin{tabular}{l} Proceedings of the CTD/WIT 2019\\ \pubnumber\\
         \pubdate  \end{tabular}}}

\newenvironment{Abstract}{\begin{quotation} \begin{center} 
             \large ABSTRACT \end{center}\bigskip 
      \begin{center}\begin{large}}{\end{large}\end{center} \end{quotation}}

\newenvironment{Presented}{\begin{quotation} \begin{center} 
             PRESENTED AT\end{center}\bigskip 
      \begin{center}\begin{large}}{\end{large}\end{center} \end{quotation}}

\def\Acknowledgements{\bigskip  \bigskip \begin{center} \begin{large}
      \bf ACKNOWLEDGEMENTS \end{large}\end{center}}

\def\pt{p_\mathrm{T}}
\def\mus{\mathrm{\mu s}}

\def\beq{\begin{equation}}
\def\eeq#1{\label{#1}\end{equation}}
\def\eeqn{\end{equation}}

\def\beqa{\begin{eqnarray}}
\def\eeqa#1{\label{#1}\end{eqnarray}}
\def\eeqan{\end{eqnarray}}

\let\bar=\overbar

\def\Dslash{\not{\hbox{\kern-4pt $D$}}}
\def\dslash{\not{\hbox{\kern-2pt $\del$}}}

\def\msb{{\bar{\ssstyle M \kern -1pt S}}}

\usepackage{color}
\usepackage{lineno}
\usepackage{subfig}
\usepackage{hyperref}

\newcommand\pubnumber{PROC-2019-027}

\newcommand\pubdate{\today}

\def\affiliation{
Department of Physics \\
Imperial College London, UK}

\newcommand{\conference}{Connecting the Dots and Workshop on Intelligent Trackers (CTD/WIT 2019)\\
Instituto de F\'isica Corpuscular (IFIC), Valencia, Spain\\ 
April 2-5, 2019}

\usepackage{fancyhdr}
\definecolor{mygrey}{RGB}{105,105,105}
\begin{document}


\large
\begin{titlepage}
\pubblock

\vfill
\Title{Level-1 Track Finding with an all-FPGA System at CMS for the HL-LHC}
\vfill

\Author{Thomas James on behalf of the CMS Collaboration}
\Address{\affiliation}
\vfill

\begin{Abstract}
The Compact Muon Solenoid (CMS) experiment at the Large Hadron Collider (LHC) is designed to study a wide range of high energy physics phenomena. It employs a large all-silicon tracker within a 3.8\,T magnetic solenoid, which allows precise measurements of transverse momentum ($p_\mathrm{T}$) and vertex position. This tracking detector will be upgraded to coincide with the operation of the High-Luminosity LHC, which will provide luminosities of up to $7.5\,\times\,10^{35}\,\mathrm{cm^{-2}s^{-1}}$ to CMS, or 200 collisions per 25\,ns bunch crossing. This new tracker must maintain the nominal physics performance in this more challenging environment. Novel tracking modules that utilise closely spaced silicon sensors to discriminate on charged particle $\pt$ have been developed and allow the selective readout of hits compatible with tracks of $\pt\,>\,2-3$\,GeV to off-detector trigger electronics. This would allow the use of tracking information at the Level-1 trigger of the experiment, a requirement to keep the Level-1 triggering rate below the 750\,kHz target, while maintaining physics sensitivity. This paper presents a concept for an all-FPGA based track finder using a fully time-multiplexed architecture. Hardware demonstrators have been assembled to prove the feasibility and capability of such a system. The performance for a variety of physics scenarios will be presented, as well as the proposed scaling of the demonstrators to the final system.\end{Abstract}

\vfill

\begin{Presented}
\conference
\end{Presented}
\vfill
\end{titlepage}
\def\thefootnote{\fnsymbol{footnote}}
\setcounter{footnote}{0}
%

\normalsize 


\section{CMS and the HL-LHC}
\label{intro}

The Compact Muon Solenoid (CMS) experiment\,\cite{cms} is a large, all purpose particle detector, designed to investigate a wide range of physics at the Large Hadron Collider (LHC)\,\cite{lhc}. CMS was designed to operate with an average number of simultaneous collisions (pileup) of $\sim 25$, with a bunch crossing rate of 40\,MHz. A major feature of CMS is the 1.2\,m radius, 200 $\mathrm{m}^2$ area silicon strip tracker, the largest silicon tracker in operation in the world. As the tracker sits within the 3.8\,T superconducting solenoid, the transverse momentum, $\pt$, of charged particles can be measured from the track curvature. 

CMS operates a Level-1 (L1) trigger to reject uninteresting events\,\cite{l1trig}. This enacts a rate reduction on the order of a factor of 400. At the time of construction, it would have been unfeasible to read-out the tracker data at 40\,MHz to help in the Level-1 decision making, primarily due to the large data size and rate.

By 2026, the LHC will be upgraded in luminosity to about $5-7.5\,\times\,{10}^{34} \mathrm{cm}^{-2}\,\mathrm{s}^{-1}$ (or $140-200$ pileup). This will enable an increase in integrated luminosity of approximately 3 times by 2035, with respect to no such upgrade. This upgraded LHC machine is known as the High-Luminosity LHC (HL-LHC)\,\cite{hllhc}, and is targeting a lifetime integrated luminosity of 3,000\,-\,4,000\,${\mathrm{fb}}^{-1}$.

During the long-shutdown preceding HL-LHC operation, large sections of the CMS detector will be replaced. The silicon tracker will have accumulated significant radiation damage, and a replacement is being constructed\,\cite{cmsph2tp, ph2trackertdr}. Simulation studies show that with this luminosity, a new handle is needed at the L1 trigger stage in order to maintain thresholds and sensitivity to interesting physics, while at the same time keeping the L1 accept rate within the expected limit of 750\,kHz. Present thresholds without track information would approach a 4\,MHz L1 accept rate at 200 pileup~\cite{cmsl1interim}. The new outer-tracker will therefore incorporate a novel design to allow the read-out of tracking information at 40\,MHz to the L1 trigger.

\section{The CMS Phase 2 Upgrade}
\label{cmsph2}

\begin{figure}[!htb]
  \centering
  \includegraphics[width=0.8\linewidth]{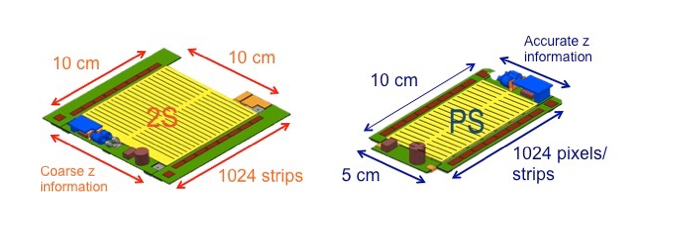}
  \caption{Layout of the two-strip (2S) and pixel-strip (PS) $\pt$ modules, as being developed for the upgraded CMS Outer Tracker. The PS modules contain a layer of 1.5\,mm macro-pixels, which give finer granularity in the direction perpendicular to the strips. These modules will be used in the regions $r\,<\,60$\,cm, where the hit occupancy is higher.}
  \label{fig:modules}
\end{figure}

New tracking modules, under development for the HL-LHC, utilise a pair of closely spaced silicon sensors ($1.6-4.0$\,mm) and on-detector correlation logic in order to discriminate charged particle tracks with a $\pt$ exceeding a threshold of $2-3$\,GeV. The module design is depicted in Figure~\ref{fig:modules}. A pair of clusters (one in each sensor) consistent with such a track is known as a `stub'. Only the stubs are sent to the off-detector L1 trigger electronics (at 40\,MHz), enacting a rate reduction of approximately a factor of ten. There remains, however, an average of 12,000\,-\,15,000 stubs per bunch crossing which must be processed, and matched to particle trajectories. 

The upgraded L1 trigger for CMS at the HL-LHC will include a `correlator' layer, which will combine information from the calorimeter, muon, and tracker systems, in order to make physics objects which can be selected on with a menu-like global trigger. The combination of tracker and calorimeter objects greatly improves the pileup rejection capabilities at L1. Particle and pileup identification algorithms have been shown to significantly improve the performance of both transverse energy ($E_\mathrm{T}$) and missing $E_\mathrm{T}$ triggers\,\cite{pfpuppi}. Vertex finding algorithms have also been implemented in FPGA logic, and have been demonstrated to perform well for physics events with high energy jets, even in the presence of $140-200$ pileup.

\begin{figure}[!htb]
  \centering
  \includegraphics[width=0.9\linewidth]{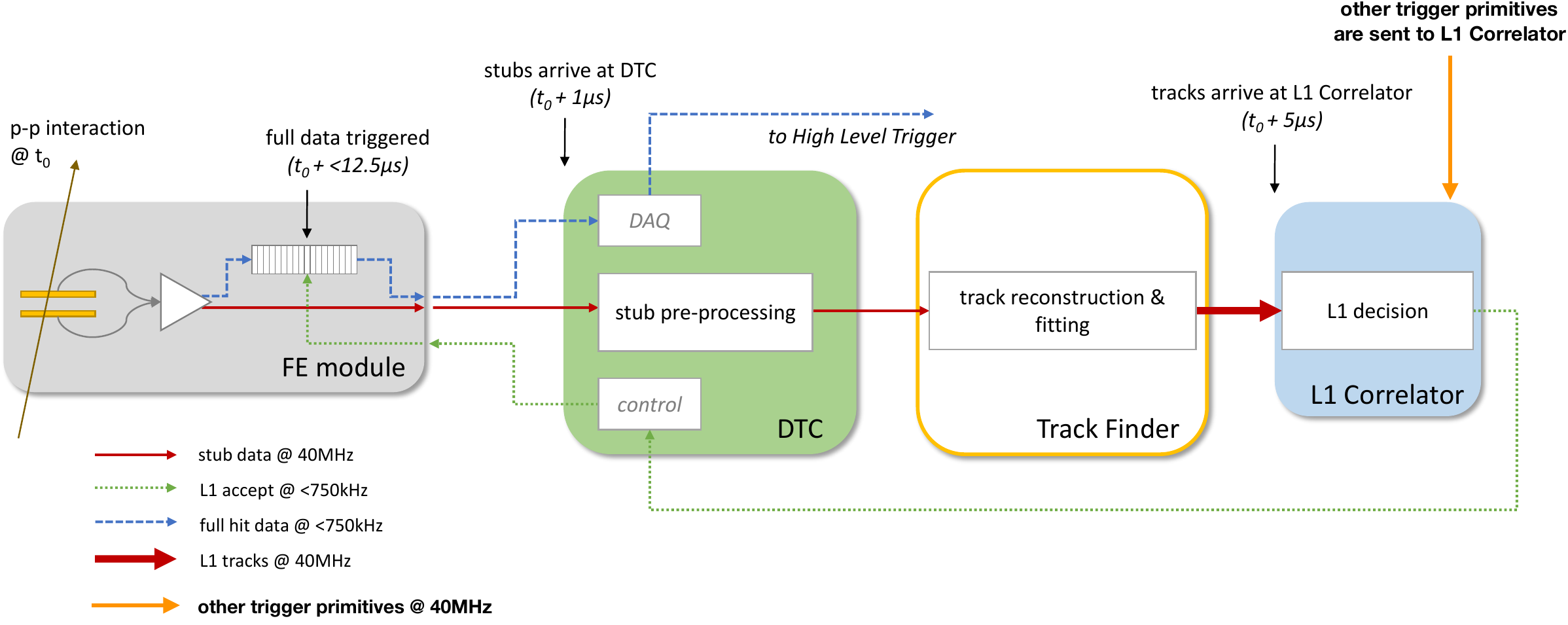}
  \caption{Illustration of dataflow and latency requirements from $\pt$ modules through to the off-detector electronics that are dedicated to forming the L1 trigger decision.}
  \label{fig:dataflow}
\end{figure}

The latency budget of the L1 trigger is 12.5\,$\mus$, determined by the depth of the front-end buffers. However, it will take an estimated 1\,$\mus$ to transmit the data between the front-ends of the detector and the back-end electronics. A further 1\,$\mus$ is required to propagate the L1 accept signal back to the front-end chips. It is estimated to take 3.5\,$\mus$ to correlate the trigger primitives from each sub-detector, and make a triggering decision. This leaves only 4\,$\mus$ for the track-finding and fitting process, accounting for a 30\% safety margin. On a L1 accept signal, all front-end buffers will be triggered to read out the information from the selected event to the data acquisition (DAQ) and High Level Trigger (HLT) farm. This architecture is illustrated in Figure~\ref{fig:dataflow}.

\section{The Back-end System}
\label{subsection}

The tracker back-end system is responsible for constructing tracks from stubs at every bunch-crossing. It must also control and receive the data from the front-end, and provide a DAQ path for the full dataset on an L1 trigger. The system is divided into two layers of FPGA-based processing: the DAQ, Trigger and Control (DTC) boards; and the Track Finder Processors (TFPs).

A total of 216 DTC boards will be used to read-out the data from the 13,296 modules of the outer-tracker. Up to 72 modules will be connected per board, with bi-directional optical links at 2.56\,Gb/s to detector, and either 5.12 or 10.24\,Gb/s from detector. The DTC boards are envisaged to host dual Xilinx Ultrascale+ FPGAs\,\cite{ultrascale}, allowing for a total bandwidth of about 0.7\,Tb/s per card. This layer of the tracker back-end is also responsible for stub pre-processing, which includes: the conversion from a local module-level coordinate scheme to a global position; the sorting of stub data into $\phi$-sectors; and the time-multiplexing of the data into one of 18 time-nodes\,\cite{tmt}.  The cabling between detector and DTC ensures that each of nine sets of 24 DTCs receives data from a single nonant (40 degree region) of the tracker in the azimuthal angle $\phi$. In this scheme, the 216 DTC boards will occupy 18 ATCA\,\cite{atca} crates in total, corresponding to two crates (one rack) per nonant.

The TFP layer of the tracker back-end accumulates data from 48 DTCs; one 25\,Gb/s optical link from each. There are expected to be 162 TFP boards in total (excluding spares/redundant nodes), each responsible for one nonant in $\phi$, and one out of every 18 bunch-crossings (the time-multiplexing period). These boards will host one or two large FPGA(s), delivering enough processing power to find and fit the track candidates. As the majority of stubs are associated with pileup, the data-rate between the TFP and the L1 correlator is reduced down to around 30\,Gb/s. 

\section{Track-finder Algorithms}
\label{algos}

Several track-finding algorithms have been studied for Level-1 tracking. Hardware demonstrators have been constructed to prove the feasibility of two candidate algorithms: a Hough Transform followed by a Kalman Filter (KF)\,\cite{tmtt, tjames}; and a `tracklet' seeding and projection followed by a $\chi^2$ minimisation fit\,\cite{tracklet}.  

A hybrid of these two algorithm options is currently under development. This solution consists of a tracklet seeding and projection, followed by a Kalman Filter. 

\begin{itemize}
    \item Tracklet seeds are formed from pairs of stubs in adjacent tracker layers and disks. Using the constraint that the track originates at the interaction point, a set of tracklet helix parameters can be calculated.
    \item The tracklet parameters are projected to the remaining layers and disks of the tracker, and stubs inside the projection windows are matched to form track candidates.
    \item Track candidates that share stubs in three or more tracker layers or disks are merged into a single candidate. This removes duplicate candidates prior to the track fit.
    \item The tracklet helix parameters are then used to seed the Kalman Filter state and covariance matrices. The matched stubs are then applied one-by-one to the KF. The KF selects the set of stubs that best fit a track, and calculates four helix parameters: $p_\mathrm{T}$, $\eta$ and $\phi$, and the longitudinal impact parameter $z_0$. 
\end{itemize}

\begin{figure}[!htb]
  \centering
  \includegraphics[angle=-90, width=\linewidth]{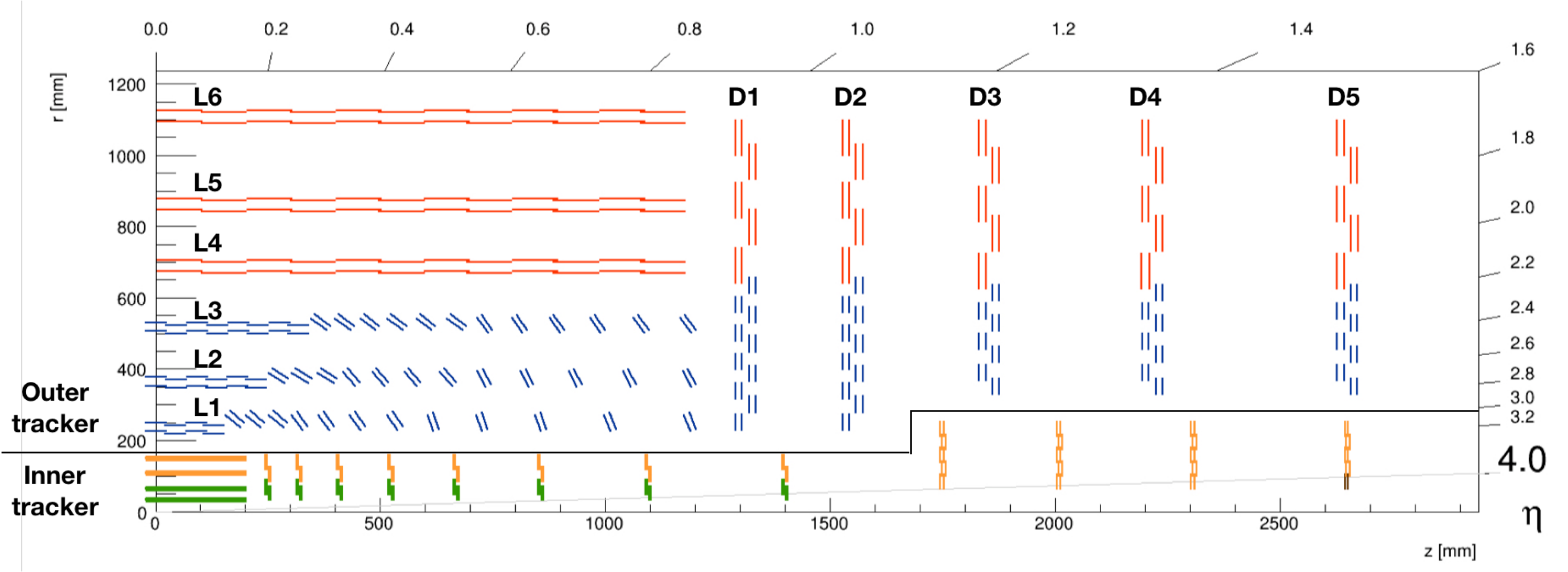}
  \caption{One quadrant of the upgraded tracker geometry and layout. The region within around 20\,cm is the Inner Tracker, and will not output data to the L1 trigger path. Six barrel layers and five endcap disks each side make up the Outer Tracker, which consist of $\pt$ modules.}
  \label{fig:layout}
\end{figure}

Seeds are generated in redundant layer and disk-pairs, to allow for inefficiencies. Figure~\ref{fig:layout} depicts one quadrant of the upgraded tracker geometry. The following pairs of layers and disks of the outer tracker are used to seed tracking (where 'L' denotes a barrel layer, and 'D' an endcap disk): L1L2, L3L4, L5L6, L1D1, L2D1, D1D2, D3D4. As the KF is able to check all possible combinations of matched stubs to build and fit the best track, wider search windows can be used in the matching stage than with the simpler $\chi^2$ minimisation approach. Stubs used to build the initial tracklet are not re-fitted by the KF, saving latency. The merging of duplicates before the fit reduces the number of track candidates that must be processed by the KF by a factor of three (from around 570 to 175 candidates). Most duplicates (71\%) are generated by the same particle being found in different seeding layers. The remainder are generated from multiple stub combinations associated with a single particle being found within the same seeding configuration. The merging operation costs about 1.5\% in track-finding efficiency as a result of nearby tracks from separate particles being combined.

In contrast to the previous demonstrator projects which were written entirely in standard Hardware Description Languages (HDLs), the 'hybrid' solution is being implemented with a High Level Synthesis (HLS) language developed by Xilinx. This change should allow for easier and faster algorithm development, quicker uptake of non-experts, and improved long-term maintainability. As the Xilinx HLS code can be run outside of an FPGA, it also acts as a software emulator for the implemented algorithms. Some aspects of the firmware design, in particular the low-level board and link infrastructure (typically developed by engineers rather than physicists) are expected to remain in HDL. The hybrid algorithm implementation is currently targeting a 250\,MHz clock frequency.

\subsection{The Kalman Filter}
\label{kf}
\begin{figure}[!htb]
\vspace{0.2pt}
  \centering
  \includegraphics[angle=-90, width=0.8\linewidth]{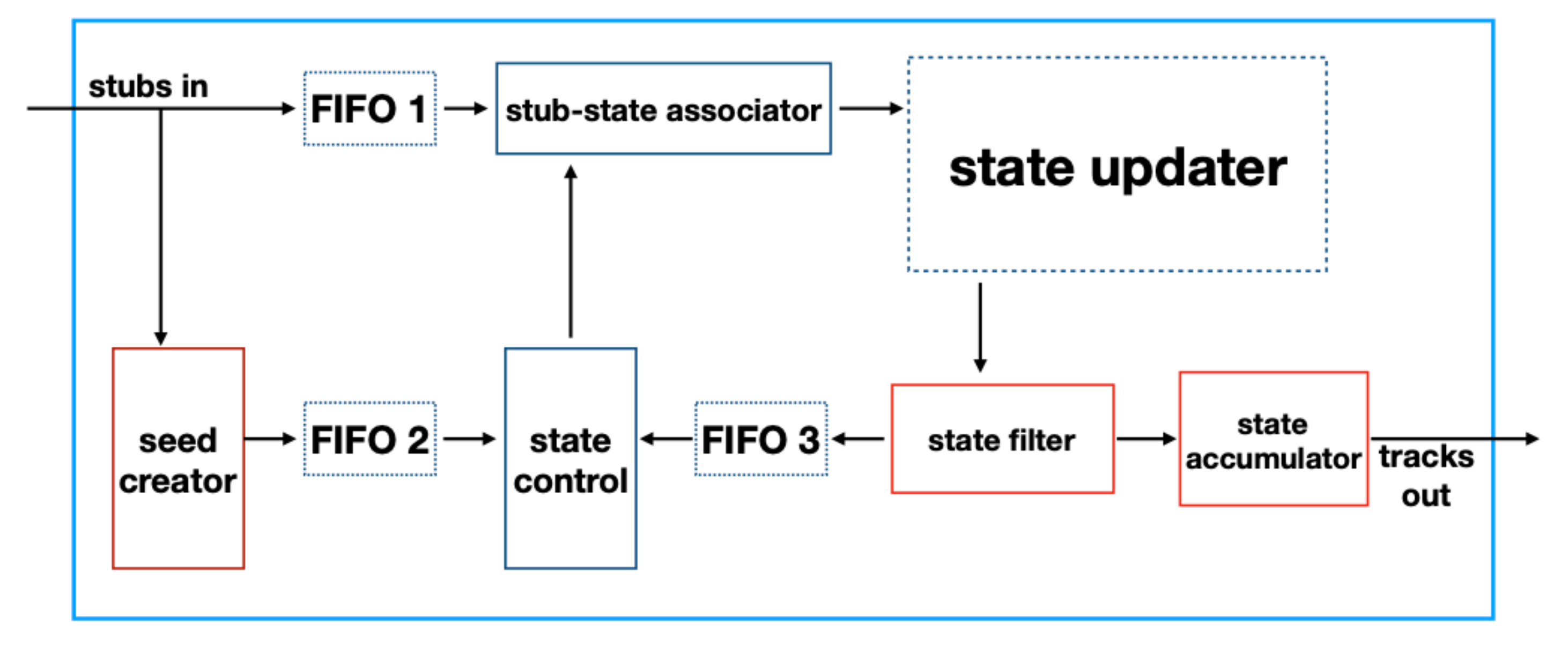}
  \caption{Simplified block diagram of a single KF worker. }
  \label{kfworker}
  \vspace{0.2pt}
\end{figure}

Each Kalman Filter worker (an independent firmware block that that runs the KF, and can process one stub per clock cycle) consists of a state updater, connected to data-flow controlling logic. A simplified block-diagram of a single worker is shown in Figure~\ref{kfworker}. Incoming stubs are stored in FIFO 1. They are later retrieved by the stub-state associator, which matches stubs to states, in order of increasing radii. The states are managed by the state control, which multiplexes partially worked states (from FIFO 3), and incoming seeds (from FIFO 2). Each state-stub combination is then passed through the state updater, which produces new state and covariance matrices from the weighted average of the seed, previous stub inputs, and new stub inputs. Finally, the most appropriate state for each input track candidate is selected, primarily based on the $\chi^2$ of the fit.

 The latency of the state updater, in three configurations of clock frequency and FPGA choice are given in Table~\ref{tab:kfstateup_lat}. The resource utilisation of a single state updater block is given in Table~\ref{tab:kfstateup_resource}. For the KU115 FPGA, the HDL control flow logic runs at 320\,MHz, and takes 18 clocks (56\,ns).

\begin{table}[!htb]
  \begin{center}
    \vspace{0.3pt}
    \begin{tabular}{l|ccc}
      \hline
      \hline
      Device &  Frequency [MHz] & Number of Clocks & Latency [ns] \\
      \hline
      KU115 & 320 & 46 & 144 \\
      KU115 & 440 & 55 & 125 \\
      VU9P & 440 & 40 & 91 \\
      \hline
      \hline
    \end{tabular}
    \caption{Latency of a single Kalman state updater, in three configurations of clock frequency and FPGA choice.}
    \label{tab:kfstateup_lat}
  \end{center}
\end{table}
\newpage
Recent improvements have been made to the Kalman Filter firmware implementation. Corrections for non-radial endcap strips improve performance at high $p_\mathrm{T}$. A simple implementation of uncertainties due to multiple scattering have been is added, whereby the hit errors in $\phi$ are inflated by 0.75\,mrad/$\pt$. The off-diagonal terms in the higher order circle expansion terms are replaced by constant shifts in $\phi$. These constants can be read from the tracklet seed, and give performance approximately equivalent to the full simulation, without significantly increasing FPGA resource utilisation. A five-parameter configuration of the fit has also been developed, which includes the transverse impact parameter ($d_0$). The KF is capable of fitting between four and six stubs per track.
 
 \begin{table}[!htb]
  \begin{center}
    \begin{tabular}{l|ccc}
      \hline
      \hline
      Object &  DSPs & BRAM (36\,Kb) \\
      \hline
      4 parameter state updater & 52 & 1.5  \\
      5 parameter state updater & 67 & 2.0  \\
      HDL control-flow & 1 & 14.5 \\
      \hline
      \hline
    \end{tabular}
    \caption{Resource utilisation of the Kalman Filter.}
    \label{tab:kfstateup_resource}
  \end{center}
\end{table}
Each Kalman Filter worker is independent, and no more than 18 are expected to be required per TFP. In total, this implementation uses 13\% BRAM, 3\% LUTs, and 17\% DSPs of a KU115 FPGA for the four parameter option, and runs at 320\,MHz. 
\section{Performance Results}
\label{results}

\begin{figure}[!htb]
  \centering
  \includegraphics[width=\linewidth]{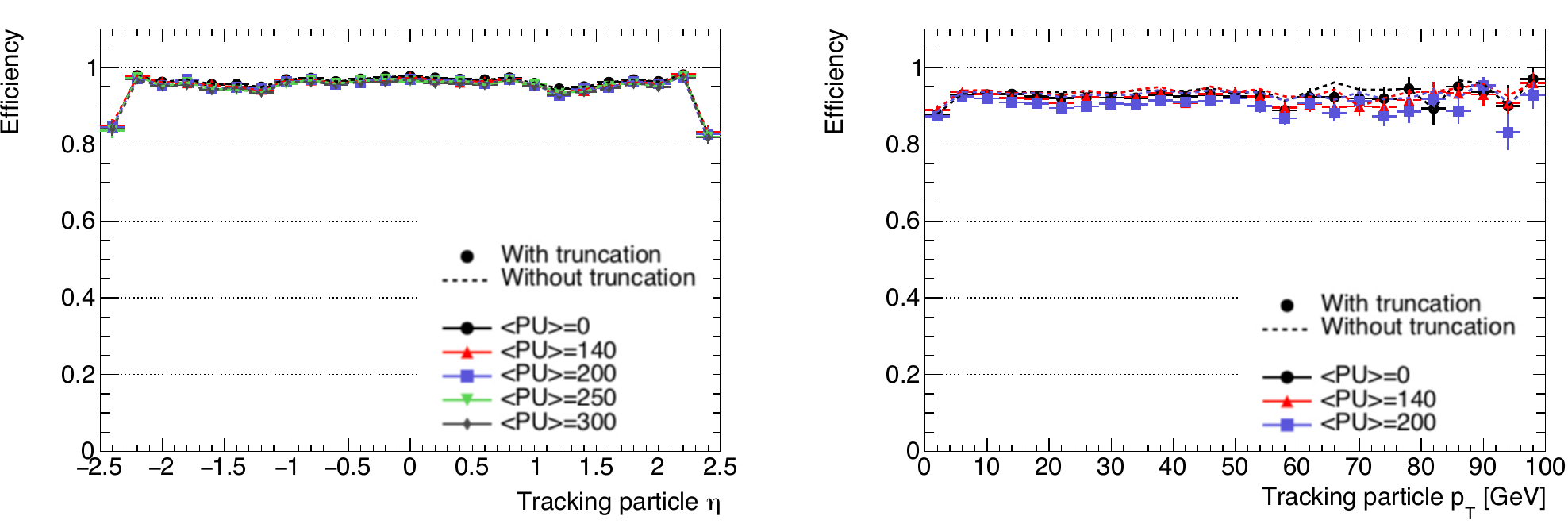}
  \caption{ Track finding efficiency against particle $\eta$ for particles with $\pt$ above 2\,GeV, produced in top quark pair production events, in simulated conditions of 0, 140, 200, 250 and 300 pileup. (left);  Track finding efficiency against particle $\pt$, for particles produced in simulated top quark pair production events, in conditions of 0, 140, and 200 pileup (right). The efficiency is shown including data loss due to the fixed latency cut-off (with truncation), and without such effects (without truncation).
}
  \label{fig:ttbar}
\end{figure}

\begin{figure}[!htb]
  \centering
  \includegraphics[width=\linewidth]{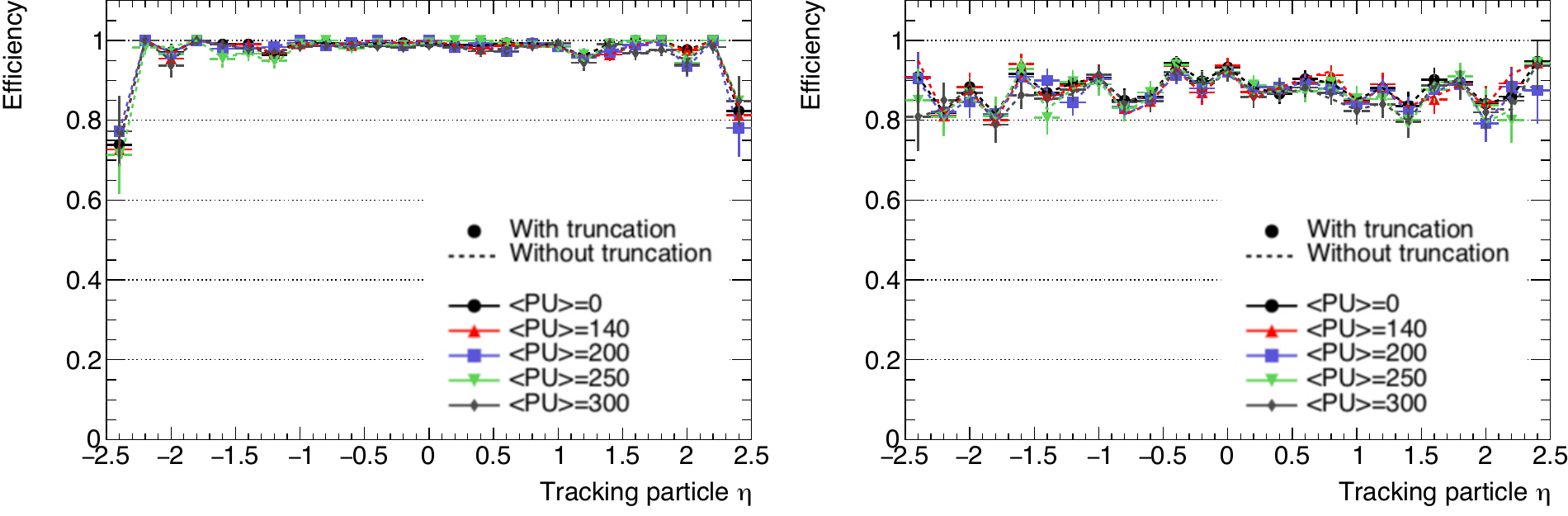}
  \caption{Track finding efficiency against particle $\eta$ for muons (left), and electrons (right). Particles with a $\pt$ above 2\,GeV are included, in simulated conditions of 0, 140, 200, 250 and 300 pileup. The efficiency is shown including data loss due to the fixed latency cut-off (with truncation), and without such effects (without truncation). 
  }
  \label{fig:lepton}
\end{figure}

\begin{figure}[!htb]
  \centering
  \includegraphics[width=\linewidth]{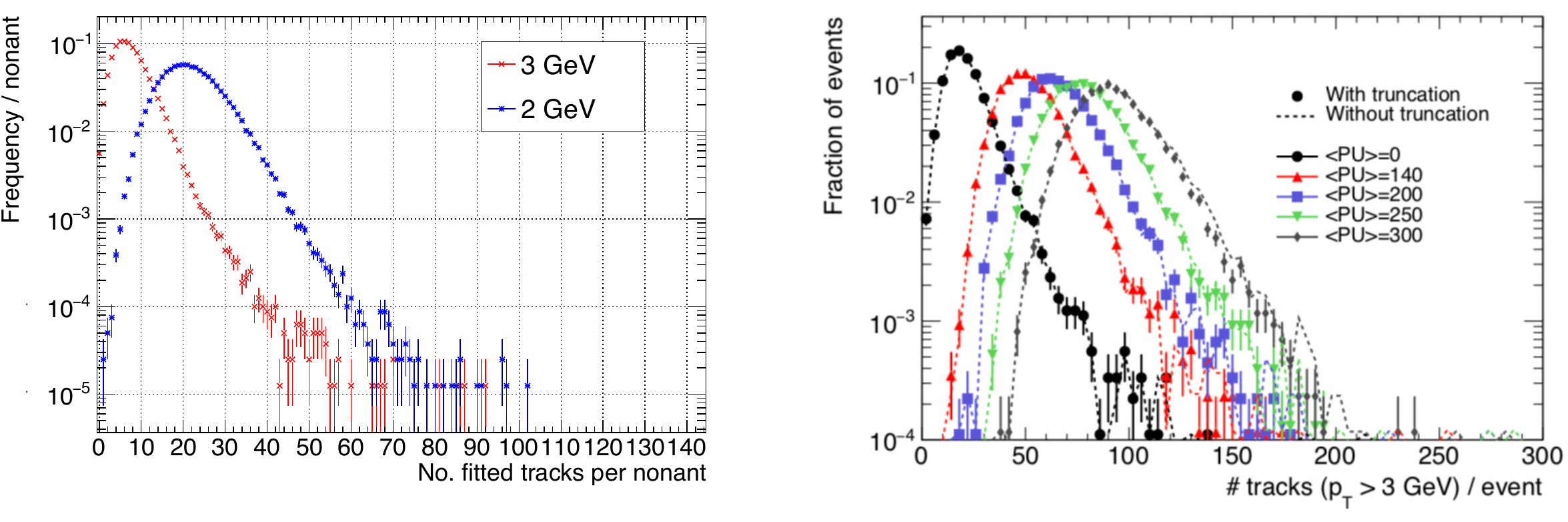}
  \caption{Track finding rates per nonant, per bunch crossing, for tracking in top quark pair production events under conditions of 200 pileup (left). Track finding rates per bunch crossing, for tracking above 3\,GeV (right) in top quark pair production events under conditions of 0, 140, 200, 250 and 300 pileup (right). The rate is shown including data loss due to the fixed latency cut-off (with truncation), and without such effects (without truncation).}
  \label{fig:rates}
\end{figure}

In simulations of top quark pairs with 200 superimposed pileup events, an average track finding efficiency of about 95\% has been accomplished for tracks with $\pt$ above 2\,GeV. This is shown in Figure~\ref{fig:ttbar}. With this same sample, an average of 60 (200) tracks are found above 3 (2)\,GeV, per bunch crossing. The distribution per bunch crossing is shown in Figure~\ref{fig:rates}. Of the tracks found, 11\% are `fake' or incorrectly reconstructed, and about 4\% are duplicates. Figure~\ref{fig:lepton} shows the tracking efficiency for leptons. The efficiency for muon track finding is in excess of 97\% above a $\pt$ of 2\,GeV. Above 10\,GeV, the efficiency for electron reconstruction is about 90\%. 

As can be seen in Figure~\ref{fig:lepton}, the track finding algorithm has been demonstrated to work well up to 300 pileup, showing significant margin for scenarios in which the HL-LHC delivers higher than expected luminosity.



\section{Displaced Track Finding}
\label{displaced}

A modification of the hybrid algorithm is being developed to allow for displaced track finding for $d_0\,<\,10$\,cm\,\cite{displaced}. As the interaction-point constraint can no longer be applied, tracklets must instead be formed from triplets of stubs. The following triplet combinations are being considered: L2L3L4, L2L3D1, L3L5L6, L2D1D2. These seeding layers would be run in parallel to those described in Section~\ref{algos}, however, in this scenario some of the prompt seeding combinations may be made redundant, and could therefore be removed with optimisation. 

A rate increase of 1.2 (1.4) times is observed when running with displaced seeding, with respect to prompt seeding only, followed by the 5 parameter (4 parameter) Kalman filter. 


\section{Conclusions}

In order to maintain physics performance under HL-LHC conditions, CMS requires tracking at Level-1 of the triggering chain. Flexible and scalable track-finder and track-fitting algorithms, running on FPGA devices have been developed, and have been operated successfully in currently available hardware. 

\newpage
\Acknowledgements
The author is supported by the Science \& Technology Facilities Council (STFC). This work was done within the context of the CMS Collaboration.



\end{document}